\documentclass{ws-ijmpa}

\usepackage{epsfig}

\begin{document}

\markboth{P. Bicudo}
{The Pentaquarks in the Linear Molecular Heptaquark Model}

%
\catchline{}{}{}{}{}
%

\title{THE PENTAQUARKS IN THE LINEAR MOLECULAR HEPTAQUARK MODEL}

\author{\footnotesize P. Bicudo}
\address{Dep. F\'{\i}sica and CFIF, Instituto Superior T\'ecnico,
Av. Rovisco Pais, 1049-001 Lisboa, Portugal,
\\ email : bicudo@ist.utl.pt}

\maketitle

\begin{abstract}
In this talk, multiquarks are studied microscopically in a standard quark model. 
In pure ground-state pentaquarks the short-range interaction is computed 
and it is shown to be repulsive. 
An additional quark-antiquark pair is then considered, and this is 
suggested to produce linear molecular system, with a narrow decay width. 
The quarks assemble in three hadronic clusters, and the central hadron provides 
stability. The possible crypto-heptaquark hadrons with exotic pentaquark 
flavours, with strange, charmed and bottomed quarks,  
are predicted. 
\end{abstract}


\par
Exotic multiquarks are expected since the early works of Jaffe
\cite{Jaffe},
and the masses and decays in the SU(3) exotic anti-decuplet were first 
predicted within the chiral soliton model
\cite{Diakonov1}. 
The pentaquarks have been revived recently by several searches of the
$\Theta^+$(1540) 
\cite{Nakano,Barmin,Stepanyan,Barth,Asratyan,Kubarovsky,Airapetian,Juengst,Aleev,Bai,Abdel-Bary,Knopfle,Aslanyan,Chekanov,Pinkenburg,Troyan,Raducci,Abt}, 
first discovered at LEPS
\cite{Nakano},
and by searches
of the $\Xi^{--}$(1860)
\cite{Alt,Fischer,Price}
and of the $D^{*-}p$(3100)
\cite{H1}, 
observed respectively at NA49
\cite{Alt}
and at H1
\cite{H1}. 
Pentaquark structures
have also been studied on the lattice
\cite{Csikor,Sasaki,Chiu1,Chiu2,Mathur2,Okiharu,Csikor2,Ishii,Alexandrou}. 
Moreover multiquarks are favoured by the presence of several 
different flavours
\cite{Richard,Cheung,Lutz}. 
The observation of the $D^{*-}p$(3100) at H1, the 
observation of double-charmed baryons at SELEX 
\cite{SELEX},
and the future search of double-charmed baryons at COMPASS
\cite{COMPASS}
suggest that new pentaquarks with one or two heavy quarks remain to be discovered. 

\par
It is well known that a narrow 
pentaquark must contain an excitation, to prevent a decay width
of hundreds of MeV to a meson-baryon channel. 
Here I consider that a s-wave flavour-singlet 
light quark-antiquark pair $l \bar l$ is added to the pentaquark $M$. The 
resulting heptaquark $M'$ is a state with parity opposite 
to the original $M$ 
\cite{Nowak}, 
due to the intrinsic parity of fermions and anti-fermions. 
The ground-state of $M'$ is also naturally rearranged in a s-wave baryon 
and in two s-wave mesons, where the two outer hadrons are 
repelled, while the central hadron provides stability.
Because the s-wave pion is the lightest hadron, the minimum energy needed to 
create a quark-antiquark pair can be as small as 100-200 MeV. This energy shift 
is lower than the typical energy of 300-600 MeV of spin-isospin or angular 
excitations in hadrons. Moreover, the heptaquarks $M'$ low-energy p-wave decay 
(after the extra quark-antiquark pair is annihilated), results in a very narrow 
decay width, consistent with the observed exotic flavour pentaquarks.

\par
Recently this principle was used to 
suggest that the $\Theta^+$(1540) is a 
$K \bullet  \pi \bullet  N$ molecule with binding energy of 30 MeV
\cite{Bicudo00,Llanes-Estrada,Kishimoto}, 
and the $\Xi^{--}$(1862) is a $\bar K \bullet  N \bullet  \bar K$ 
molecule with a binding energy of 60 MeV
\cite{Bicudo00,Bicudo02}.
I also suggest that the new positive parity  
scalar  $D_s(2320)$ and axial $D_{s+}(2460)$ are
$\bar K \bullet  D$ and $\bar K \bullet D^*$ multiquarks 
\cite{Bicudo01},
and that the $D^{*-} p$(3100) is consistent with a  
$D^* \bullet  \pi \bullet  N$ 
linear molecule with an energy of 15 MeV above threshold
\cite{Bicudo00,Bicudo03}.
I now predict new exotic strange, charmed and bottomed 
pentaquarks compatible with the linear molecular heptaquark model.


%
%
\begin{table}[t]
\tbl{ 
Exotic-flavour pentaquarks with no heavy quark.
}
{\begin{tabular}{@{}cccc@{}}
\toprule
flavour & linear molecule   
& $~$ mass [GeV] & decay channels \\ \colrule
$ I=1/2, \ ssss \bar l (+3 \, l \bar l) $ : &five-hadron molecule
\\
\colrule
$ I=1, \ sssl \bar l  (+2 \, l \bar l) $ : & four-hadron molecule
\\
\colrule
$ I=3/2, \ ssll \bar l  (+l \bar l)  = $ & $s \bar l \bullet   lll \bullet   s \bar l :$ 
\\
& $ \bar K \bullet   N \bullet   \bar K {\bf = \Xi^{--} } $&  {\em 1.86 } &$ 
\bar K + \Sigma, \, \pi + \Xi
$\\
\colrule
$I=2, \ slll \bar l  (+l \bar l) = $ & $s\bar l \bullet   lll \bullet   l \bar l $: &pion unbound
\\
\colrule
$I=5/2, \ llll \bar l (+l \bar l) = $ & $l\bar l \bullet   lll \bullet   l \bar l$: & pion unbound
\\
\colrule
$I=0, \ llll \bar s (+l \bar l) = $ & $l \bar s \bullet   l\bar l \bullet   lll $ :
\\   
& $K \bullet   \pi \bullet   N { \bf =\Theta^+ }  $& {\em 1.54 }  &$ 
K + N 
$\\ 
\botrule
\end{tabular}}
\end{table}

In this talk, multiquarks are studied
microscopically in a standard quark-model (QM) Hamiltonian. 
The energy of the multiquark state, and the short range interaction of 
the mesonic or baryonic subclusters of the multiquark are both computed 
with the multiquark matrix element of the QM Hamiltonian,
\begin{equation}
H= \sum_i T_i + \sum_{i<j} V_{ij} +\sum_{i \bar j} A_{i \bar j} \ . 
\label{Hamiltonian}
\end{equation}
Each quark or antiquark has a kinetic energy $T_i$.
The colour-dependent two-body interaction $V_{ij}$ includes 
the standard QM confining and hyperfine terms,
\begin{equation}
V_{ij}= \frac{-3}{16} \vec \lambda_i  \cdot   \vec \lambda_j
\left[V_{conf}(r) + V_{hyp} (r) { \vec S_i } \cdot { \vec S_j }
\right] \ .
\label{potential}
\end{equation}
The potential of eq. (\ref{potential}) 
reproduces the meson and baryon spectrum with quark and antiquark
bound states (from heavy quarkonium to the light pion mass).
Moreover, the Resonating Group Method (RGM)
\cite{Wheeler}
was applied by Ribeiro 
\cite{Ribeiro} 
to show that in exotic
$N + N$ scattering the quark two-body-potential, together with
the Pauli repulsion of quarks, explains the $N + N$ hard core
repulsion. 
Recently, addressing a tetraquark system with $\pi+\pi$ quantum 
numbers, it was shown that the QM with the quark-antiquark 
annihilation $A_{i \bar j}$ also fully complies with chiral 
symmetry, including the Adler zero and the Weinberg theorem
\cite{Bicudo0,Bicudo1,Bicudo2}. 

\par
For the purpose of this talk, only the matrix elements of the
potentials in eq. (\ref{Hamiltonian}) matter. 
The hadron spectrum constrains the hyperfine potential
\begin{equation}
\langle V_{hyp} \rangle \simeq \frac{4}{3} 
\left( M_\Delta-M_N \right)
\simeq M_{K^*}- M_K  \ .
\label{hyperfine}
\end{equation}
From the pion mass
\cite{Bicudo3},
the matrix elements of the annihilation potential are,
\begin{equation}
\langle A \rangle_{S=0} \simeq - {2 \over 3} 
(2M_N-M_\Delta)
\ ,
\label{sum rules}
\end{equation}
which is correct for the annihilation of $u$ or $d$ quarks.

%
\begin{table}[t]
\tbl{
Exotic flavour pentaquarks with one heavy quark. 
}
{\begin{tabular}{@{}cccc@{}}
\toprule
flavour& linear molecule   
& $~$ mass [GeV] & decay channels \\ \colrule
$ I=1/2, Hsss \bar l (+2 \, l \bar l) $ : & four-hadron molecule
\\ 
\colrule
$ I=1, Hssl \bar l (+l \bar l)= $ & $     s \bar l \bullet  l l H \bullet  s \bar l   :$ 
\\
&$ \bar K \bullet  \Lambda_c \bullet  \bar K $&$  3.23 \pm 0.03 $&$  \bar K + \Xi_c  , \, \pi +\Omega_c 
$\\
&$ \bar K \bullet  \Lambda_b \bullet  \bar K $&$  6.57 \pm 0.03 $&$  \bar K + \Xi_b  , \, \pi +\Omega_b
$\\
\colrule
$ I=3/2, Hsll \bar l (+l \bar l)=  $ & $    s \bar l \bullet  lll \bullet  H \bar l   :$
\\
&$  \bar K \bullet  N \bullet  D $&$ 3.25 \pm 0.03 $&$ \bar K + \Sigma_c  , \, D + \Sigma , \, \pi + \Xi_c 
$\\ 
&$ \bar K \bullet  N \bullet  D^* $&$  3.39 \pm 0.03 $&$  \bar K + \Sigma_c  , \, D^* + \Sigma , \, \pi + \Xi_c 
$\\ 
&$ \bar K \bullet  N \bullet  \bar B $&$ 6.66 \pm 0.03 $&$ \bar K + \Sigma_b  , \, \bar B + \Sigma , \, \pi + \Xi_b

$\\ 
&$ \bar K \bullet  N \bullet  \bar B^* $&$  6.71 \pm 0.03 $&$  \bar K + \Sigma_b  , \, \bar B^* + \Sigma , \, \pi + \Xi_b

$\\
\colrule
$ I=2, Hlll \bar l (+l \bar l)= $ & $ l \bar l \bullet  l l l \bullet  H \bar l   $ 
:  & pion unbound
\\
\colrule  
$ I=1/2, Hlll \bar s  (+l \bar l)= $ & $ l \bar s \bullet  l \bar l \bullet  l l H  :$ 
\\  
&$ K \bullet  \pi \bullet  \Sigma_c $&$ 3.08 \pm 0.03 $&$  K + \Lambda_c , \, K + \Sigma_c  , \, D_s +N
$\\
&$ K \bullet  \pi \bullet  \Sigma_b $&$ 6.41 \pm 0.1  $&$  K + \Lambda_b , \, K + \Sigma_b  , \, D_s +N
$\\
$ I=1/2, Hlll \bar s  (+l \bar l) = $ & $ l \bar s \bullet  H \bar l \bullet  lll  :$ 
\\  
&$ K \bullet  \bar D \bullet  N $&$  3.25 \pm 0.03 $&$  K + \Lambda_c , \ K + \Sigma_c , \ D_s + N 
$\\
&$ K \bullet  \bar D^* \bullet  N $&$ 3.39  \pm 0.03 $&$  K + \Lambda_c  , \ K + \Sigma_c  , \ D_s^* + N 
$\\
&$ K \bullet  \bar B \bullet  N $&$  6.66 \pm 0.03 $&$  K + \Lambda_b  , \ K + \Sigma_b , \ B_s + N  
$\\
&$ K \bullet  \bar B^* \bullet  N $&$ 6.71 \pm 0.03 $&$   K + \Lambda_b  , \ K + \Sigma_b , \ B_s^* + N 
$\\
\botrule
\end{tabular}}
\end{table}

\par
The annihilation potential only shows up in non-exotic channels, and it 
is clear from eq. (\ref{sum rules}) that the annihilation potential 
provides an attractive (negative) interaction. 
The quark-quark(antiquark) potential is 
dominated by the interplay of the hyperfine interaction of eq. (\ref{hyperfine})
and the Pauli quark exchange.
In s-wave systems with low spin this results in a repulsive interaction. 
Therefore, I arrive at the attraction/repulsion criterion for groundstate hadrons:
\\
- {\em whenever the two interacting hadrons have quarks (or antiquarks)
with a common flavour, the repulsion is increased by the Pauli principle;
\\
- when the two interacting hadrons have a quark and an
antiquark with the same flavour, the attraction 
is enhanced by the quark-antiquark annihilation}.
\\
For instance, $uud-s \bar u$ is attractive,
and $uud-u \bar s$ is repulsive. 
This qualitative rule is confirmed by quantitative 
computations of the short-range interactions of the 
$\pi , \, N , \, K , \, D , \, D^* , \, B , \, B^* $ 
\cite{Bicudo00,Bicudo01,Bicudo03,Bicudo02,Bicudo0,Bicudo1,Bicudo2}.

\begin{table}[t]
\tbl{
Exotic flavour pentaquarks with one heavy anti-quark.
}
{\begin{tabular}{@{}cccc@{}}
\toprule
flavour& linear molecule   
& $~$ mass [GeV] & decay channels \\ \colrule
$ I=0, ssss \bar H (+3 l \bar l)$ : & five-hadron molecule
\\
\colrule
$ I=1/2, sssl \bar H (+2 \, l \bar l)  $ : & four-hadron molecule
\\
\colrule
$ I=0, ssll \bar H (+l \bar l) = $ & $ l \bar H \bullet  l \bar l \bullet  lss $ 
\\
&$  \bar D \bullet  \pi \bullet  \Xi $&$  3.31 \pm 0.03 $&$  \bar D + \Xi  
, \, \bar D_s + \Lambda 
$\\
&$  \bar D^* \bullet  \pi \bullet  \Xi $&$  3.45 \pm 0.03 $&$  \bar D^* + \Xi
, \, \bar D^*_s + \Lambda , \, \bar D_s + \Lambda 
$\\
&$  B \bullet  \pi \bullet  \Xi $&$  6.73 \pm 0.03 $&$  B + \Xi  
, \, B_s + \Lambda
$\\
&$  B^* \bullet  \pi \bullet  \Xi $&$  6.77 \pm 0.03 $&$  B^* + \Xi  
, \, B^*_s + \Lambda , \, B_s + \Lambda
$\\
\colrule
$ I=1/2, slll \bar H (+l \bar l) = $ & $ l \bar H \bullet  l \bar l \bullet  l l s  $ 
\\
&$ \bar D \bullet  \pi \bullet  \Sigma $&$  3.19 \pm 0.03 $&$  \bar D + \Lambda , \, \bar D + \Sigma  , \, \bar D_s +N
$\\
&$ \bar D^* \bullet  \pi \bullet  \Sigma $&$  3.33 \pm 0.03 $&$  \bar D^* + \Lambda , \, \bar D^* + \Sigma  , \, \bar D^*_s +N
$\\
&$ B \bullet  \pi \bullet  \Sigma $&$  6.60 \pm 0.03 $&$  B + \Lambda , \, B + \Sigma  , \, B_s +N 
$\\
&$ B^* \bullet  \pi \bullet  \Sigma $&$  6.64 \pm 0.03 $&$  B^* + \Lambda , \, B^* + \Sigma  , \, B^*_s +N
$\\
$ I=1/2, slll \bar H (+l \bar l) =   $ & $ l \bar H \bullet  s\bar l \bullet  l l l   $ 
\\
&$  \bar D \bullet  \bar K \bullet  N $&$  3.25 \pm 0.03 $&$  \bar D + \Lambda , \, \bar D + \Sigma  , \, \bar D_s +N
$\\
&$ \bar D^* \bullet  \bar K \bullet  N $&$ 3.39 \pm 0.03 $&$   \bar D^* + \Lambda , \, \bar D^* + \Sigma  , \, \bar D^*_s +N
$\\
&$ B \bullet  \bar K \bullet  N $&$ 6.66 \pm 0.03 $&$  B + \Lambda , \, B + \Sigma  , \, B_s +N
$\\
&$ B^* \bullet  \bar K \bullet  N $&$  6.71 \pm 0.03 $&$  B^* + \Lambda , \, B^* + \Sigma  , \, B^*_s +N
$\\
\colrule
$ I=0, llll \bar H (+l \bar l)= $ & $ l \bar H \bullet  l \bar l \bullet  lll $
\\
& $ \bar D \bullet  \pi \bullet  N  $&$  2.93 \pm 0.03 $&$ \bar D+ N  
$\\ 
& $ \bar D^* \bullet  \pi \bullet  N {\bf  = \bar D^{*-}p } $& {\em 3.10 } &$ \bar D^*+ N , \, \bar D +N
$\\ 
& $ B \bullet  \pi \bullet  N $&$ 6.35 \pm 0.03 $&$ B+N  
$\\ 
& $ B^* \bullet  \pi \bullet  N $&$  6.39 \pm 0.03 $&$ B^*+N, B+N 
$\\ 
\botrule
\end{tabular}}
\end{table}


Again, the attraction/repulsion criterion shows that the exotic pentaquarks 
containing five quarks only are repelled. To increase binding we include 
a light $l \bar l$ quark-antiquark pair in the system. 
I now detail the strategy to find the possible linear heptaquark molecules. 
\\
{\bf a)} The top quark is excluded because it is too unstable. 
To minimise the short-range repulsion and to increase the attraction of 
the three-hadron system, I only consider pentaquarks with a minimally
exotic isospin, and with low spin. 
\\
{\bf b)} Here the flavour is decomposed in an s-wave system of 
a spin $1/2$ baryon and two 
pseudoscalar mesons, except for the vectors $D^*$ and $B^*$ 
which are also considered. 
\\
{\bf c)} I consider as candidates for narrow pentaquarks
systems where one hadron is attracted by both other ones.
The criterion is used to discriminate which hadrons
are bound and which are repelled. 
\\
{\bf d)} In the case of some exotic flavour pentaquarks, only a 
four-hadron-molecule or a five-hadron-molecule would bind. These cases 
are not detailed, because they are difficult to create in the laboratory.
\\
{\bf e)} Moreover, in the particular case where one of the three hadrons 
is a $\pi$, binding is only assumed if the $\pi$ is the central hadron, 
attracted both by the other two ones. The $\pi$ is too light 
to be bound by just one hadron
\cite{Bicudo00}.
\\
{\bf f)} The masses of the bound states with a pion are computed assuming a total 
binding energy of the order of 10 MeV, averaging the binding energy of the
$\Theta^+$ and of the $D^{*-}p$ system in the molecular perspective. The masses of the 
other bound states are computed assuming a total binding energy of the order of 50 MeV, 
averaging the binding energies of the $\Xi^{--}$ and of the new positive-parity $D_S$
mesons. 
\\
{\bf g)}
Here higher excitations are neglected (they would further increase the binding energy). 
This results in an error bar of $\pm$ 30 MeV for the mass.
When one of the hadrons in the molecule is not listed by the Particle Data Group
\cite{RPP}, 
its mass is extracted from a recent lattice computation
\cite{Mathur},
and the error bar is $\pm$ 100 MeV.
\\
{\bf e)}
Although three-body decay channels are possible through quark rearrangement,
their observation requires high experimental statistics. Only some of the 
different possible two-body decay processes are detailed here.


\par
To conclude, this work has performed a systematic search of exotic-flavour pentaquarks, 
using the heptaquark, or linear three-body hadronic-molecule perspective. 
This perspective is the result of standard QM computations
of pentaquarks and hepatquark masses and of hadron-hadron short-range interactions. 
A large number of new exotic flavour-pentaquarks are predicted in Tables
1, 2 and 3
together with their two-body decay channels. 
The systems with more than one heavy antiquark are very numerous and they are omitted 
here. It is interesting to remark that degenerate states are shared by Tables 2 and 3.
Moreover, some new multiquarks may be easier to bind than the presently observed 
exotic pentaquarks.

\section*{Acknowledgments}
I am grateful to Katerina Lipka, Achim Geiser, Paula Bordalo and Pedro Abreu for 
discussions on the possibility to detect new exotic pentaquarks. This talk is 
devoted to encourage the experimental search for new multiquarks.


\end{document}